\documentclass{iaus}
\usepackage{graphicx}

\def\spose#1{\hbox to 0pt{#1\hss}}
\newcommand\lsim{\mathrel{\spose{\lower 3pt\hbox{$\mathchar"218$}}
     \raise 2.0pt\hbox{$\mathchar"13C$}}}
\newcommand\gsim{\mathrel{\spose{\lower 3pt\hbox{$\mathchar"218$}}
     \raise 2.0pt\hbox{$\mathchar"13E$}}}

\title[JD 11.~~Cosmic Chemical Evolution with Brightest Events] 
{The Cosmic Chemical Evolution as seen by the Brightest Events in the Universe}

\author[Sandra Savaglio]   
{Sandra Savaglio$^1$}

\affiliation{$^1$Max Planck Institute for Extraterrestrial Physics,\\ 85748 Garching bei M{\"u}nchen, Germany \\ email:{\it savaglio@mpe.mpg.de}}

\pubyear{2009}
\volume{265}  
\pagerange{119--126}
\jname{Chemical Abundances in the Universe: Connecting First Stars to Planets}
\editors{K. Cunha, M. Spite \& B. Barbuy, eds.}
\begin{document}

\maketitle

\begin{abstract}
Gamma-ray bursts (GRBs) are the brightest events in the universe. They have been used in the last five years to  study  the cosmic chemical evolution, from the local universe to the first stars.  The sample  size is still relatively small when compared to field galaxy  surveys. However, GRBs show a universe that is surprising. At $z>2$, the cold interstellar medium in galaxies is chemically evolved, with a mean metallicity of about 1/10 solar. At lower redshift ($z<1$), metallicities of the ionized gas are relatively low, on average 1/6 solar. Not only is there no evidence of redshift evolution in the interval $0<z<6.3$, but also the dispersion in the $\sim30$ objects is large. This suggests that the metallicity of host galaxies is not the physical quantity triggering GRB events. From the investigation of other galaxy parameters, it emerges that active star-formation might be a stronger requirement to produce a GRB.  Several recent striking results strongly support the idea that GRB studies open a new view on our understanding of galaxy formation and evolution, back to the very primordial universe at $z\sim8$.

\end{abstract}

\firstsection 
\section{Introduction}

During the last decade, the chemical evolution of the universe has been investigated using a new class of objects: gamma-ray bursts (GRBs). GRBs are the brightest sources in the universe, but were first detected only in 1967 by a US military satellite (Klebesadel et al.\ 1973), because their emission does not last long. For this reason, their cosmological origin was demonstrated only in 1997, when the first redshift was measured (Metzger et al.\ 1998). Today, after more than twelve years, the number of events with spectroscopic redshift is still relatively low, about 200. Nevertheless, on April 23 2009 the highest spectroscopic redshift ever was measured, and this happened to be a GRB, GRB~090423, at $z=8.2$ (\cite[Salvaterra et al.\ 2009]{salvaterra09}; Tanvir et al.\ 2009). This is not only a very exciting success for GRB science, but it also demonstrates that the field is still potentially and effectively crucial for the understanding of our universe. 

GRBs are very luminous, but do not shine for very long (it cannot be any different, otherwise we would not be here to tell). Their $\gamma$-ray emission lasts at most a few minutes, during which they radiate the same energy emitted by the Sun over its entire life, 10 Gyr. Long-duration GRBs (more than a few seconds, the majority of those detected) originate from the final core collapse of a massive star, a supernova (Woosley, 1993). Short-duration GRBs (shorter than a few seconds) have  likely a different progenitor (Katz  \& Canel 1996): the coalescence of two compact objects (neutron stars or black holes). In both classes, rotation is the key ingredient producing the collimated emission. The GRB rate is of the order of one event every $10^5$ years in a galaxy. This means that, integrating over the entire universe and considering the collimated emission, few events are detectable from $\gamma$-ray satellites. 

During the last two years, a number of particularly interesting discoveries have shown that the universe probed by GRBs is surprisingly exciting. Apart from the already mentioned GRB~090423, in March 2008 the brightest source ever was recorded. This was GRB~080319 at $z=1.9$ (7.5 Gyrs after the Big Bang), nicknamed the ``naked eye''  GRB because it had an optical magnitude $m=5.6$ at its maximum (Bloom et al.\ 2009). In September 2008, the at-the-time second most distant object ever was detected, GRB~080913B at $z=6.7$ (Greiner et al.\ 2009).

Thanks to this rich phenomenology and the large redshift range spanned, there is no doubt that GRBs are very effectively probing, among other things, the chemical evolution of the universe, all the way from the local universe to the epoch of first stars, more than 13 Gyr ago. In this paper, we will summarize the results obtained in the last five years.

\begin{figure}[b]
\begin{center}
 \includegraphics[width=4.5in]{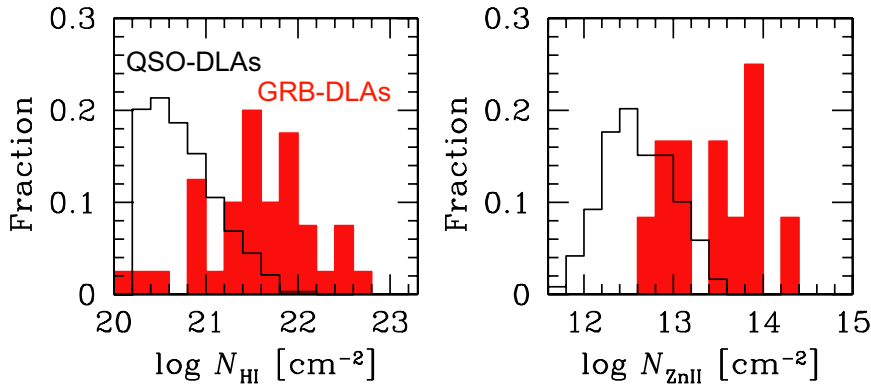} 
 \caption{Fraction of GRB-DLAs (filled histograms) and QSO-DLAs (empty histograms) per HI and ZnII column-density bin (left- and right-hand side panels, respectively). The QSO-DLA histograms are complete for
$\log N_{\rm HI} > 20.2$ and $\log N_{\rm ZnII} > 12.4$. The completeness level for GRB-DLAs is not well determined. It is apparent that column densities in GRB-DLAs are generally higher than in QSO-DLAs. This can either indicate that GRB-DLAs originate in bigger galaxies, or that the volume density of the gas is higher (e.g., the GRB sightline is crossing a region closer to the galaxy center) than QSO-DLAs, or both.}
   \label{f1}
\end{center}
\end{figure}

\begin{figure}[b]
\begin{center}
 \includegraphics[width=4in]{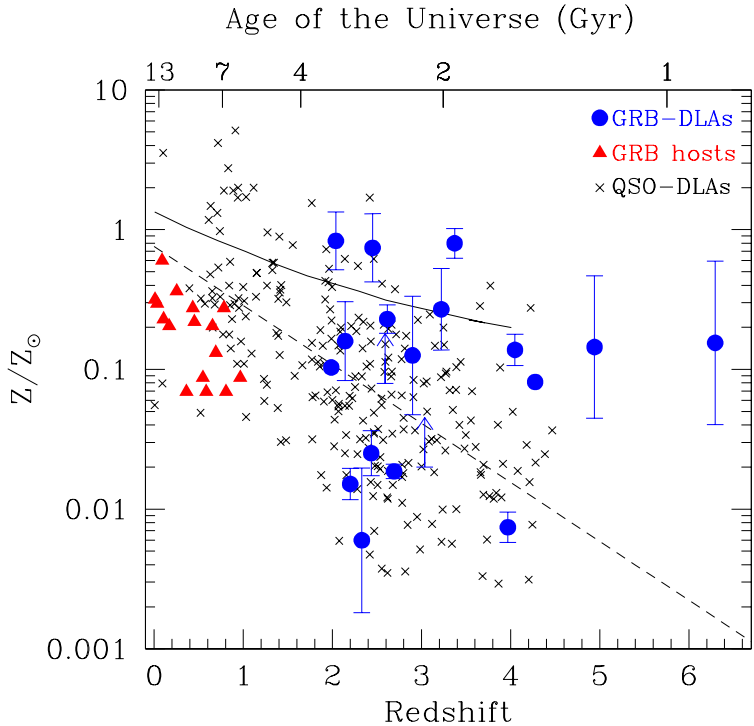} 
 \caption{Redshift evolution of the metallicity relative to solar values, for 17 GRB-DLAs at $z>2$, 16 GRB hosts at $z<1$  and $\sim250$ QSO-DLAs in the interval $0<z<4.4$. Error bars are not available for all GRB-DLAs. Errors for GRB hosts are not estimated. Errors for QSO-DLAs are generally smaller than 0.2 dex. The dashed line is the best-fit linear correlation for QSO-DLAs. The solid line is the mean metallicity predicted by semi-analytic models for galaxy formation (Somerville et al.\ 2001). The GRB-DLAs metallicity in $2<z<4.5$ is on average 2.5 times higher than the average value in QSO-DLAs in the same redshift interval.}
   \label{f2}
\end{center}
\end{figure}

\section{The cosmic chemical enrichment with GRBs}

There are basically two distinct methods providing information on the chemical enrichment in galaxies and its redshift evolution using GRBs. In one case, rest-frame UV absorption lines detected in the optical afterglow spectra give measurements in the neutral gas ($T\lsim1000$ K) for $z>2$ host galaxies (e.g., Savaglio et al.\ 2003; Prochaska et al.\ 2007; Fynbo et al.\ 2009). In the other, rest-frame optical emission lines from integrated spectra of $z<1$ hosts  probe the ionized gas ($T\gsim 5000$ K; e.g., Soderberg et al.\ 2004; Gorosabel et al.\ 2005; Th\"one 2008). These two complementary methods did not give so far results simultaneously for the same GRB event for lack of suitable instrumentation. However, the newly commissioned optical-NIR spectrograph X-Shooter at the ESO Very Large Telescope and the Cosmic Origin Spectrograph recently installed on Hubble Space Telescope should fill the {\it redshift desert} soon. The former already delivered interesting findings for the $z=3.372$ event GRB~090313 (de Ugarte Postigo et al.\ 2009).

The absorption systems seen in high-$z$ GRBs are called GRB-DLAs, as they are similar to damped Lyman-$\alpha$ systems (DLAs) detected in QSO spectra. One difference is that in the former case, the DLA is in the host galaxy, while in the latter case the DLA is generally not associated with the QSO and distributed along its sight line. Moreover, column densities in GRB-DLAs are generally higher than in QSO-DLAs (Fig.~\ref{f1}), indicating that the neutral-gas regions crossed by GRBs are larger, or denser, or both, than those crossed by QSOs. In Fig.~\ref{f2} we show the metal abundances measured in the DLAs detected in GRB hosts and in QSO sight lines. It was claimed, from a smaller sample, that GRB-DLAs have generally higher metallicity than QSO-DLAs (Berger et al.\ 2006; Savaglio 2006; Prochaska et al.\ 2007). The most up-to-date sample of GRB-DLAs, shown in Fig.~\ref{f2}, contains 17 measurements and two lower limits in the redshift interval $2<z<6.3$. The average value (and statistical dispersion) for the 15 GRB-DLAs in  $2.0 < z < 4.5$ is $<$${\rm [Z/H]}$$>$ $=-1.0\pm0.7$, whereas for the 156 QSO-DLAs in the same redshift interval this is $<$${\rm [Z/H]}$$>$ $=-1.4\pm0.6$. The new large sample of GRB-DLAs tends to show still a higher metal content than QSO-DLAs, but the gap is getting smaller. This indicates that the observational bias that prevents us from measuring abundances when metal lines are too weak might affect our results. The difference with QSO-DLAs is that GRB afterglows, when spectroscopically observed, are on average several magnitudes fainter than the typical QSO, and they cannot be observed for too long because they disappear quickly.

For lower redshift, $z<1$, metallicities are measured with emission lines from HII regions in the host galaxy. Emission line metallicities rely on different calibrators (Kewley \& Ellison 2008) used depending on the set of lines available, according to the GRB redshift and instrument setting. Results on less than 20 hosts indicate metallicities between solar and 1/14 times solar values (Savaglio, Glazebrook \& Le Borgne 2009; Levesque et al. 2009). The average value and dispersion in 16 hosts (median redshift $z=0.44$) is  $<$${\rm [Z/H]}$$>$ $=-0.75\pm0.29$ (Fig.~2;  Savaglio et al.\ 2009). This is somehow surprising, as we do not see evidence of redshift evolution from GRB-DLA metallicities at $z>2$. On the other hand, evolution is observed in QSO-DLAs, where metallicity at $z<1$ is $<$${\rm [Z/H]}$$>$ $=-0.3\pm0.5$, a factor of at least 10 times higher than at $z>2$.

It was recently proposed that the different metallicities in GRB-DLAs and QSO-DLAs could be due to the different regions probed by the two populations. GRBs tend to occur in regions with high star-formation, therefore in regions closer to the galaxy center, where metallicity is on average larger than in a random galaxy sightline. QSOs are background sources not associated with the galaxy hosting the DLA, therefore their sightline is crossing the galaxy in a random location, not necessarily close to a region of star formation (Fynbo et al.\ 2008). This is confirmed by the larger dust content and extinction measured in GRB-DLAs with respect to QSO-DLAs (Kann et al.\ 2006; Kr{\"u}hler et al.\ 2008; Prochaska et al.\ 2009). However, such a sensible conclusion collides with the relatively low metallicities found in low-$z$ GRB hosts. The large dispersion of the metal content in GRB hosts in a large redshift interval is indicative that perhaps metallicity is not driving the GRB phenomenon. For this reason, we consider in the following sections the other two fundamental physical quantities characterizing  galaxies: the star-formation rate (SFR) and the stellar mass $M_\ast$. 

\begin{figure}[b]
\begin{center}
 \includegraphics[width=4.5in]{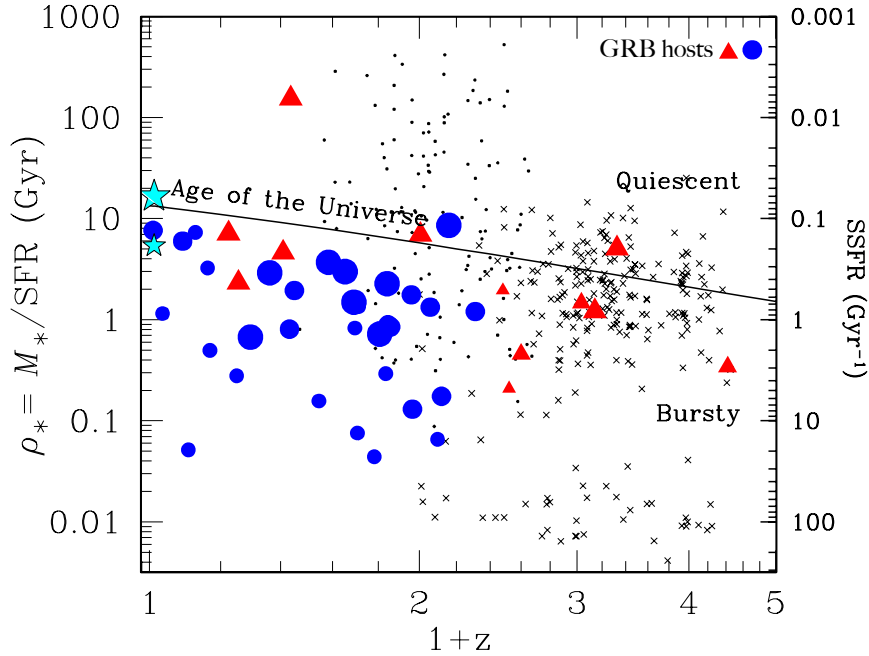} 
 \caption{Growth timescale $\rho_\ast = M_\ast/{\rm SFR}$ (left y-axis) or its inverse, the specific star-formation rate ${\rm SSFR = SFR}/M_\ast$ (right y-axis) as a function of redshift (Savaglio et al.\ 2009). Filled circles and triangles are GRB hosts with SFRs measured from emission lines and UV luminosities, respectively. Small, medium, and large symbols are hosts with $M_\ast \leq 10^{9.0}$ M$_\odot$, $10^{9.0}$ M$_\odot$ $< M_\ast \leq 10^{9.7}$ M$_\odot$, and $M_\ast > 10^{9.7}$ M$_\odot$, respectively. The curve shows the age of the universe as a function of redshift, and indicates the transition from bursty to quiescent mode for galaxies. Dots are field galaxies at $0.5 < z < 1.7$ (Juneau et al.\ 2005). Crosses are Lyman break galaxies at $1.3<z<3$ (Reddy et al.\ 2006). The big and small stars at zero redshift represent the growth timescale for the Milky Way and the Large Magellanic Cloud, respectively.}
   \label{f3}
\end{center}
\end{figure}

\section{Star-formation rate and stellar mass of GRB hosts}

Fruchter et al.\ (2006) found that most GRBs occur preferentially in the brightest regions of galaxies. This is similar to supernovae of type Ic (Kelly et al.\ 2007). GRB hosts have very often some sign of star formation. SFRs are measured from optical nebular emission lines (generally H$\alpha$ and [OII])  up to redshift $z=1.4$. For higher redshift, when nebular emission lines are redshifted to the more difficult NIR, the rest-frame UV emission (observed optical) can be used as the easiest (but more uncertain) star-formation indicator. Values measured in the interval $0<z<3.4$ span a large range, from $\sim 0.01$ M$_\odot$ yr$^{-1}$ to $\sim 40$ M$_\odot$ yr$^{-1}$ (Savaglio et al.\ 2009). The average value of 2.5 M$_\odot$ yr$^{-1}$ is relatively high, five times higher than in the Large Magellanic Cloud. However, real SFRs might be affected by dust obscuration in large portions of the host. SFRs from submillimeter fluxes (not affected by dust) in four $z\sim 1$ GRB hosts are much higher, $\sim150$ $M_\odot$ yr$^{-1}$, more than an order of magnitude larger than the optical/UV values (Micha{\l}owski et al.\ 2008). The effect of undetected dust extinction is still not totally understood.

The stellar mass of GRB hosts is derived by fitting the observed spectral energy distribution (SED) over a large wavelength range, which should include detections in the rest-frame NIR, beyond the 4000\AA\ Balmer break (Micha{\l}owski et al.\ 2008; Savaglio et al.\ 2009). This is because the bulk of the stellar mass is in small and cold stars which are mostly emitting in the NIR. The sample for which the stellar mass is determined is still relatively low, 45 objects in the redshift interval $0<z<3.4$ (average redshift $z=0.96$); the majority of them are at $z<2$ (Savaglio et al.\ 2009). On average the stellar mass is low, of the order of the stellar mass of the Large Magellanic Cloud: $M_\ast = 10^{9.3}$ M$_\odot$. 

From the observed stellar mass and metallicities, one can ask whether GRB hosts behave like normal field galaxies. In particular, one can consider the mass-metallicity (MZ) or luminosity-metallicity relations for galaxies, as a function of redshift. So far, all attempts trying to identify the two relations in GRB hosts and similarities with field galaxies have failed (e.g.\ Berger et al.\ 2007; Chen et al.\ 2009; Levesque et al.\ 2009; Savaglio et al.\ 2009). The main problem seems to be the small number statistics. 

We can compare median values  of stellar mass and metallicity for $z<1$ GRB hosts with the MZ relations for field galaxies by Tremonti et al.\ (2004) at $z\sim0.07$ and Savaglio et al.\ (2005) at $z\sim0.7$. At $z\leq0.45$ (9 GRB hosts, median redshift $z=0.17$) the median metallicity and stellar mass are $\log Z/Z_\odot = -0.56$ and $M_\ast = 10^{9.21}$ M$_\odot$, respectively. In the interval $0.55 \leq z\leq 0.97$ (7 GRB hosts, median redshift $z=0.69$) the median metallicity and stellar mass are $\log Z/Z_\odot = -1.06$ and $M_\ast = 10^{9.73}$ M$_\odot$, respectively. 

To compare these values with field-galaxy relations, we convert the MZ relations using the newly published converters of metallicity calibrators (Kewley \& Ellison 2008).  The expected metallicities in the two redshift and stellar mass bins ($z=0.17,0.69$ and $M_\ast = 10^{9.21}, 10^{9.73}$ M$_\odot$) are higher than in GRB hosts, in both cases $\log Z/Z_\odot = -0.16$. The difference is significant, especially for the high-redshift bin. This issue needs further investigations with more objects, especially at higher redshifts, where the MZ relation shows a strong redshift evolution (Erb et al.\ 206; Maiolino et al. 2007). X-Shooter is the best instrument available at this time to measure metallicity in high-$z$ GRB hosts.

From the mass and the SFR, it is possible to derive a meaningful galaxy physical parameter: the specific star-formation rate SSFR $={\rm SFR}/M_\ast$, that is the SFR per unit stellar mass. Its inverse, the growth time-scale $\rho_\ast = M_\ast/{\rm SFR}$, gives the time interval required to a galaxy to reach the observed stellar mass, assuming that the measured SFR is constant over its past history. Fig.~\ref{f3} shows $\rho_\ast$ and SSFR as a function of redshift for GRB hosts, and the comparison with field galaxies. GRB hosts are almost all star-forming galaxies, half of them are in the bursty regime.

\begin{figure}[b]
\begin{center}
 \includegraphics[width=4.5in]{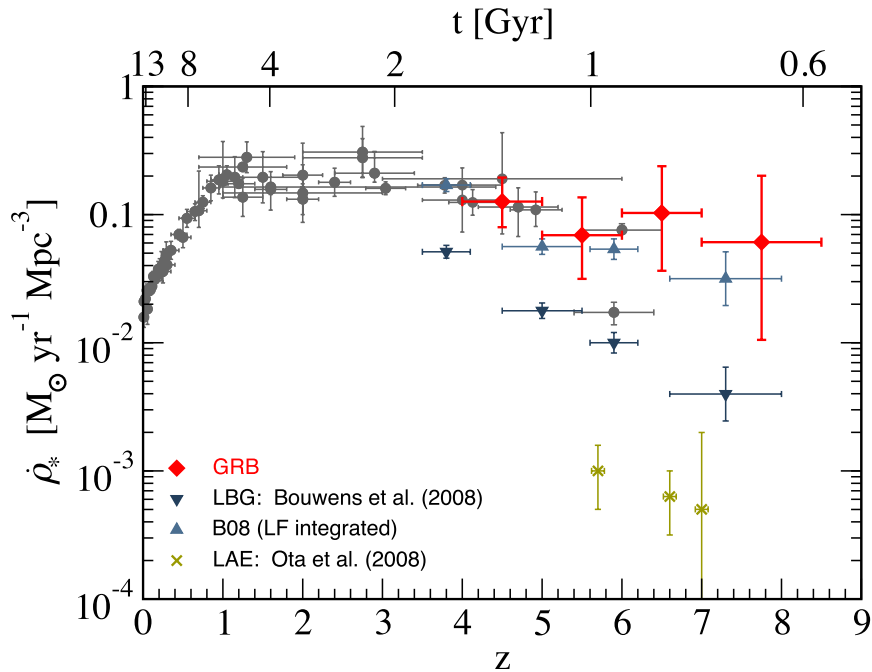} 
 \caption{The cosmic star formation density of the universe (from Kistler et al.\ 2009). Light circles are the data from Hopkins \& Beacom (2006). Crosses are contributions from Lyman-$\alpha$ emitters (LAEs; Ota et al.\ 2008). Down and up triangles are Lyman-break galaxies (LBGs) for two UV luminosity functions: integration down to 0.2$L_\ast$ at $z=3$ (Bouwens et al.\ 2008)  and complete (up triangles), respectively. The latter shows a better match with values inferred from GRBs (red diamonds; Kistler et al. 2009), indicating the strong contribution from small galaxies generally not accounted for in the observed LBG luminosity function.}
   \label{f4}
\end{center}
\end{figure}

\section{Star formation history of the universe with GRBs}

As most GRBs are associated with massive stars, therefore regions of star formation,  they are interesting candidates to study the SFR density (SFRD) of the universe. This exercise, recently attempted by Chary, Berger, \& Cowie (2007), is based on the idea that the GRB rate in galaxies at different epochs is proportional to the SFR and that the ratio does not change with redshift. The normalization is done by taking the SFR density value at low redshift for which the density of the GRB rate is estimated. 

Kistler et al.\ (2009) have compared SFRD for different field galaxy samples with SFRD derived from GRBs (Fig.~\ref{f4}).  The recent GRB~080913 at $z=6.7$ and GRB~090423 at $z=8.2$ have further extended the redshift interval where this can be done, in a regime never explored before. At $z = 8$, GRB SFRD is consistent with Lyman-break galaxy (LBG) measurements after accounting for unseen galaxies at the faint-end UV luminosity function. This implies that not all star-forming galaxies at these redshifts are currently being accounted for in deep surveys. GRBs provide the contribution to the SFRD from small galaxies. An interesting implication is that the typical GRB host at high redshift might be a small star forming galaxy. This is not totally obvious, because it has been established that the SFRD in massive galaxies was much higher in the past than it is today (at $z\sim2$ a factor of  6 higher than at $z\sim0$), whereas the redshift evolution has been milder for low-mass galaxies (Juneau et al.\ 2005). At $z>5$, we might expect SFR to be mainly in massive galaxies. Finally, the SFRD from GRBs does not show a clear decline for $z>5$, and it is much higher than in Lyman-$\alpha$ emitters (LAEs).

\section{Conclusions}

It is well known that GRBs shine through a universe that is hard to see in other ways. They are incredibly bright and last only a short time. These two properties make them very different from QSOs, which are not as bright, and do not fade away, making the investigation of nearby galaxies much more complicated. From luminous GRB afterglows, it is possible to measure the redshift and localize faint galaxies. 

Most GRBs are associated with the death of a massive star, thus with a star-forming region. It is well known that the SFR of the universe was much higher in the past than it is today (Hopkins \& Beacom 2006), therefore GRBs might be the most efficient way of identifying the evolution of the SFR density. We also know that the SFRD is dominated by small star-forming galaxies, which are probably the most common galaxies in the distant universe (Pozzetti et al.\ 2009). 

Identification of distant galaxies with GRBs is affected by a different bias than traditional galaxy surveys, because GRBs are not detected through optical instruments, but with $\gamma$-ray and X-ray satellites. Some of these galaxies can be faint, because dust extinguished, or because too far, or because intrinsically faint. With GRBs it is possible to explore extreme regimes of galaxy parameters, thus they are important to understand galaxy formation and evolution. GRB hosts identified in the optical and NIR at  $z<2$ are generally small (on average the stellar mass of the Large Magellanic Cloud) and star-forming galaxies, although SFRs span a large interval.

GRBs are probes of the state of the chemical enrichment of the universe, from the local universe, back to the time of the formation of first stars. Metallicities of the cold ISM in host galaxies at $z > 2$ is not low. The measured average value is 1/10 solar, and the dispersion is large, about a factor of five. Relatively high metallicity is confirmed also for the highest redshift detections (Savaglio 2006; Totani et al.\ 2006; Price et al.\ 2007), which means that there is no indication of redshift evolution. Low metallicities of the GRB progenitor are theoretically predicted (no mass loss) in order to keep a high angular momentum, and have a highly collimated jet.

Relatively high metallicities, in this case from ionized gas of the host galaxies, are confirmed also at $z<1$. The average value is 1/6 solar, with a dispersion of a factor of two, indicating that the metal content in host galaxies is not evolving so fast.  The sample is not very large and systematic uncertainties are still not totally under control, therefore more observations and detections are very important. Nevertheless, the lack of evidence of redshift evolution and the observed large dispersion suggest that GRBs do not happen necessarily in metal poor galaxies. Star formation, on the other hand, might be a more important physical trigger. 

GRBs are extremely important for our understanding of the primordial universe and the formation and evolution of heavy elements. The enlightening discoveries of the last few years are a clear indication that the investigation is affected by our technical capabilities which have dramatically improved recently. Dedicated instruments and observational programs have opened a new window in the hidden universe and show that this is more surprising and fascinating than expected.

\end{document}